\documentclass[aps,prl,twocolumn,superscriptaddress,showpacs,floatfix,reprint]{revtex4-1}
\usepackage{makeidx}
\usepackage{amsfonts}
\usepackage{amssymb}
\usepackage{enumerate}
\usepackage{latexsym}
\usepackage{bbm}
\usepackage{graphicx}
\usepackage{float}
\usepackage{natbib}

\newcommand{\beq}{\begin{equation}}
\newcommand{\eeq}{\end{equation}}
\newcommand{\beqa}{\begin{eqnarray}}
\newcommand{\eeqa}{\end{eqnarray}}

\newcommand{\ket}[1]{\left\vert #1 \right\rangle}

\begin{document}

\title{30 years of squeezed light generation}

\author{Ulrik L. Andersen}
\affiliation{Department of Physics, Technical University of Denmark, Fysikvej, DK-2800 Kongens Lyngby, Denmark}
\affiliation{Max-Planck-Institut for the Science of Light, G\"unther-Scharowsky-Stra\ss e 1,
Bau 24, 91058 Erlangen, Germany}
\author{Tobias Gehring}
\affiliation{Department of Physics, Technical University of Denmark, Fysikvej, DK-2800 Kongens Lyngby, Denmark}
\author{Christoph Marquardt}
\affiliation{Department of Physics, Technical University of Denmark, Fysikvej, DK-2800 Kongens Lyngby, Denmark}
\affiliation{Max-Planck-Institut for the Science of Light, G\"unther-Scharowsky-Stra\ss e 1,
Bau 24, 91058 Erlangen, Germany}
\affiliation{Institute for Optics, Information and Photonics, University Erlangen-N\"urnberg, Staudtstr. 7/B2, 90158 Erlangen, Germany}
\author{Gerd Leuchs}
\affiliation{Max-Planck-Institut for the Science of Light, G\"unther-Scharowsky-Stra\ss e 1,
Bau 24, 91058 Erlangen, Germany}
\affiliation{Institute for Optics, Information and Photonics, University Erlangen-N\"urnberg, Staudtstr. 7/B2, 90158 Erlangen, Germany}
\date{\today}

\begin{abstract}
Squeezed light generation has come of age. Significant advances on squeezed light generation has been made over the last 30 years -- from the initial, conceptual experiment in 1985 till todays top-tuned, application-oriented setups. Here we review the main experimental platforms for generating quadrature squeezed light that has been investigated the last 30 years. 
\end{abstract}

\maketitle

\section{Introduction}

The year of light 2015 marks the 1000th anniversary of the seven volume treatise on optics ``Kitab al-Manazir'' written by the scientist Ibn al-Haytham and it marks the 150th anniversary of Maxwell's equations.  
It is also a special year for the experimental quantum optics community: 2015 is the year in which we celebrate the 30th anniversary of the first generation of squeezed light. 

At the beginning of the eighties there was already an enormous literature on squeezed light on the theory side. Up to that time the experimental efforts had been in vain. To illustrate this we would like to cite from the talk Marc D. Levenson gave at the seventh International Laser Spectroscopy Conference (ICOLS VII) on Maui in the summer of 1985 after years and years of working on the topic: ``(Squeezed) states have eluded experimental demonstration, at least so far. From an experimentalist's point of view squeezed state research can be best described as a series of difficulties that must somehow be overcome''. What follows in the proceedings are nine sections, titled ``First Difficulty'' all the way up to ``Ninth Difficulty'', nothing more nothing less~\cite{Levenson1985}. 

Then in the Fall 1985, the first signature of squeezed light was observed in a groundbreaking experiment by Slusher, Hollberg, Yurke, Mertz and Valley~\cite{Slusher1985} using the process of four-wave-mixing in an atomic vapor of Sodium atoms. Despite the fierce competition between a number of groups in the USA using different technological platforms, the group of Slusher et al.\ won the squeezing race and witnessed the long-sought-after effect of squeezing -- a true quantum effect of light.  

\begin{figure*}
	\centering
		\includegraphics[scale=1]{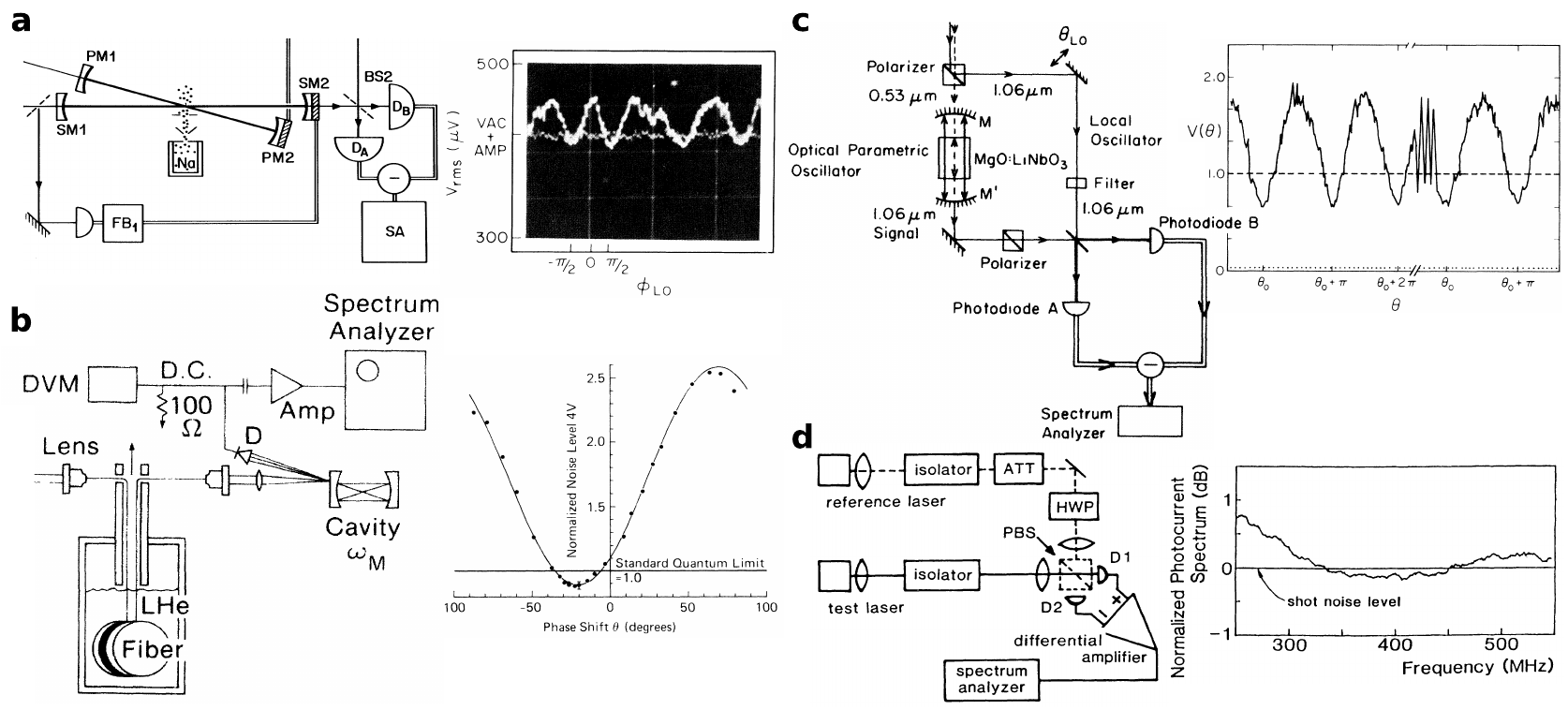}
		\caption{The figure summarizes the first three squeezing experiments.  (a) Slusher et al.\ generated squeezing in an atomic Na beam. The Na atoms were pumped in the cavity formed by the mirrors PM1 and PM2. Squeezing was generated by four-wave-mixing in the pumped Na atoms inside the standing-wave cavity formed by the mirrors SM1 and SM2. The measured squeezing was $0.3$\,dB below the vacuum noise in the squeezed quadrature, see figure on the right. The vacuum noise reference is shown in the figure by the dim trace labelled ``VAC+AMP''. (b) Shelby et al.\ also generated squeezing by four-wave-mixing, but within a $114$\,m long optical fiber coiled up and cooled to $4.2$\,K in liquid Helium. The squeezing was measured by self-homodyne detection where the phase between the local oscillator and the squeezed beam was shifted by a single-ended cavity. The measured squeezing shown on the right was about $0.6$\,dB below the vacuum noise. DVM: Digital voltmeter. (c) Wu et al.\ used parametric down-conversion in a Magnesium doped Lithium Niobate crystal embedded in a standing-wave cavity and pumped at the second harmonic of the degenerate signal and idler fields. The results are displayed on the right, where the root-mean-square noise voltage measured by the spectrum analyzer is shown versus the phase of the local oscillator. The squeezing was about $3.5$\,dB below the vacuum noise (dashed line). (d) Machida et al.\ suppressed the photocurrents driving a semiconductor laser in order to produce squeezing. This was done in the "test laser" while the "reference laser" was used as a mean to calibrated the shot noise level. The spectrum of the output of the "test laser" is shown. A maximum squeezing of 0.2dB was measured.}
	\label{first-experiments}
\end{figure*}

In strong competition with the atomic vapor technology for generating squeezed light via four-wave-mixing in 1985 was the fiber based approach exploiting the third-order Kerr type nonlinearity of $SiO_2$ as well as the approach based on the second-order nonlinearity of a ferroelectric crystal. These two technologies finally succeeded in generating squeezed light in the spring and summer of 1986~\cite{Wu1986,Shelby1986}. Soon thereafter, in December 1986, another technology for squeezed light production was presented. Based on a current noise suppression technique, Machida et al.\ managed to observe squeezing in the output of a diode laser~\cite{Machida1987}. The four experiments are presented in Fig.~\ref{first-experiments}.  

\begin{figure}
	\centering
		\includegraphics[scale=1]{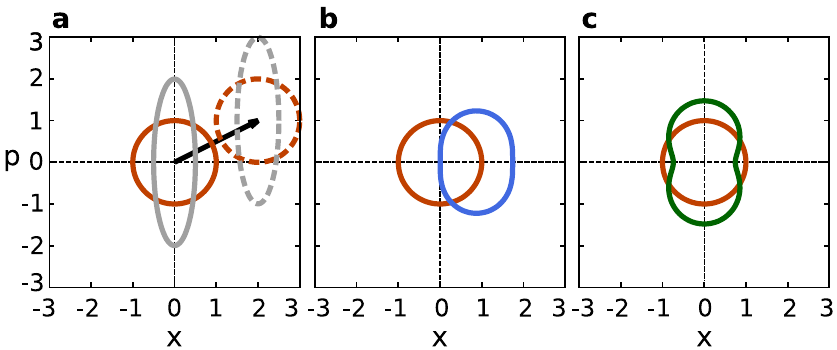}
    \caption{Phase space diagrams of (a) a vacuum (brown, solid), a $6$\,dB squeezed vacuum (grey, solid), a coherent (brown, dashed) and a $6$\,dB squeezed
coherent state (grey, dashed); (b) $\ket{\Phi} = a\ket{0} + b\ket{1}$ with $a \approx 0.87$ and $b \approx 0.5$ (blue); (c) $\ket{\Phi} = a\ket{0} + b\ket{2}$ with $a \approx 0.95$ and $b \approx -0.30$ (green). The states are shown by their standard deviation in the respective quadrature.} 
    \label{Phasespace}
\end{figure}

But what is a squeezed state? Consider first the wave function of an optical state in the position and momentum representations, $\phi(x)$ and $\phi(p)$, where $x$ and $p$ correspond to the amplitude and phase quadratures of light. The norm squared of these wave functions, $|\phi(x)|^2$ and $|\phi(p)|^2$, are the marginals of the state's Wigner function and represent the probability distributions for the amplitude and phase quadrature outcomes. For the vacuum and the coherent state, the two wave functions are rotationally symmetric Gaussians with identical widths which means that the variance associated with the measurement of any quadrature will be all the same and often normalized to unity: $V(x)=V(p)=1$. These states can be portrayed in phase space by depicting the cross sections of their respective Wigner functions as shown in Fig.~\ref{Phasespace}. A quantum state is called squeezed if the variance of a quadrature amplitude is below the variance of a vacuum or a coherent state (e.g. $V(x)<1$). This comes at the expense of having the conjugated quadrature variance being above the variance of the vacuum (e.g. $V(p)>1$) in order to obey Heisenberg's uncertainty relation. Typical examples of squeezed states are the squeezed vacuum and squeezed coherent states as shown in Fig.~\ref{Phasespace}a. These states remain Gaussian as the squeezing transformation is a Gaussian map. However, squeezed states can be also non-Gaussian -- one example is the simple superposition of vacuum, $|0\rangle$, and a single photon state, $|1\rangle$: $|\Phi\rangle=a|0\rangle+b|1\rangle$ which is squeezed by 1.3\,dB below the vacuum noise limit \cite{Wodkiewicz1987} or the two-photon state $|2\rangle$: $|\Phi\rangle=a|0\rangle+b|2\rangle$ which is squeezed by 2.6\,dB. They are illustrated in Fig.~\ref{Phasespace}b and~\ref{Phasespace}c, respectively.

The original definition of squeezed states refers to the squeezing of the quadrature amplitudes but squeezing of other quantities has also been studied in the literature. This includes squeezing of the photon numbers -- known as photon number squeezing -- where the photon number distribution is squeezed below the distribution of a coherent state, as well as the squeezing of the polarization Stokes parameters which is a form of two-mode squeezing~\cite{Caves1985,Reid1988} reminiscent of quadrature squeezing for bright beams while similar to photon number squeezing for dim beams~\cite{Muller2015}.     

The first encounter of squeezed states in the literature (although not coined squeezed state at that time) appeared in 1927 in a paper by Earle Kennard~\cite{Kennard1927}. He treated the evolution of a generic Gaussian wave packet of a harmonic oscillator under the constraints of the Heisenberg uncertainty relation by which squeezed states were born. The theory of Kennard was finalized in Copenhagen where he had discussions with Heisenberg and Bohr who are acknowledged in his paper. Many years later the theory was formalized by introducing the famous squeezing operator~\cite{Plebanski1956}, and further generalizations were made by Takahashi~\cite{Takahashi1965}, and Miller and Miskin~\cite{Miller1966}. A detailed review of squeezed states was presented by Yuen who proposed to coin it ``two-photon coherent states''~\cite{Yuen1976} closely related to the ``contractive states'' of a mechanical system~\cite{Yuen1983}. Hollenhorst talked about the "squeeze" operator~\cite{Hollenhorst1979} while Caves finally suggested the name that is used today: ``squeezed states''~\cite{Caves1981}. A number of review papers on squeezed states have been written over the years~\cite{Walls1983,Leuchs1986,Leuchs1988,Teich1989,Dodonov2002,Reid2009,Lvovsky2015,Chekhova2015}, but most of these papers are devoted to the theoretical treatment of squeezed states. In the present Review, we will focus on experimental developments in generating squeezed light during the last 30 years.    
 
Since the first milestone-marking experiments on squeezed light generation in 1985 and 1986, a number of other groups world-wide have embarked on route towards generating stronger squeezing using different technologies. The squeezing technology has been continuously improved with low-loss optical components, high-efficiency detectors and low-noise electronics leading to very large squeezing degrees of $9$\,dB using atomic ensembles, $7$\,dB using optical fibers and $13$\,dB using ferroelectric crystals. This constitutes significant improvements compared to the initial experiments where $0.3$\,dB, $0.6$\,dB and $3.5$\,dB, respectively, were observed in these systems. On the occasion of the 30th anniversary of the first squeezed light experiment, we will review the main achievements on generating squeezed light during the last three decades starting from the initial pioneering experiments to the current state-of-the-art.

\section{Squeezed light from parametric down-conversion}

Parametric down-conversion (PDC) has a long-standing reputation as a generator of squeezed light. It was first experimentally realized by Wu et al.\ in 1986~\cite{Wu1986} -- just one year after the first demonstration of squeezed light. It was a scientific breakthrough as this new source of squeezed light exhibited a significant improvement in the generation of squeezed light (3.5\,dB squeezing) compared to the atomic squeezing experiment from 1985 where 0.3\,dB squeezing was observed.

In PDC, a pump photon with frequency $\omega_p$, incident on a dielectric with a $\chi^{(2)}$ nonlinearity, breaks up into two new photons; a signal photon of frequency $\omega_s$ and an idler photon of frequency $\omega_i$ where $\omega_p=\omega_i+\omega_s$. For degenerate PDC, the idler and signal photons are indistinguishable in frequency ($\omega_i=\omega_s$) and polarization; otherwise the process is non-degenerate. 

The $\chi^{(2)}$ nonlinearity is in general very weak and in order to observe a significant induced polarization of the medium and hence efficient squeezing, different considerations have to be taken into account. First, the concentrated field in the crystal has to be relatively high which is either solved by using high power pulsed lasers or by placing the crystal inside a cavity. Experiments using these two approaches are discussed in the following sections. Second, the momentum of the involved fields has to be conserved, that is $k_p=k_s+k_i$ where $k_p$ is the wave vector for the incident pump beam while $k_s$ and $k_i$ are the wave vectors associated with the signal and idler beams, respectively. This is the phase matching condition, and can be achieved by temperature and wavelength tuning and/or periodically poling the non-linear crystal.  

In the frequency and polarization degenerate case, corresponding to Type I phase matching, the system Hamiltonian can be written in the form 
\begin{eqnarray}
H=\kappa(a^2-a^{\dagger 2})\ ,
\end{eqnarray}
where $a$ is the annihilation operator for the signal (or idler) field, and $\kappa$ is the non-linear coupling parameter. The unitary evolution of the input signal under this Hamiltonian is $U=\exp(-i\kappa(a^2-a^{\dagger 2}))$ which is exactly the form of the so-called squeezing operator.   

\subsection{Optical parametric oscillation}

An optical parametric oscillator (OPO), where the parametric down-conversion process takes place inside a cavity, has proven to be the most efficient source of quadrature squeezed light. In most cavity configurations, the signal and idler modes are resonant thereby enhancing the effective non-linearity but also enriching the dynamics. E.g.\ for degenerate PDC, the cavity is introducing a critical condition (a pitch-fork bifurcation) in the system which is the well-known threshold for optical parametric oscillation. For an ideal system without losses, infinite squeezing is expected around this threshold point. If the degenerate signal and idler modes of the OPO are seeded with a bright beam, the bifurcation point disappears and the OPO works as a squeezing amplifier also known as a phase-sensitive amplifier.  

The experiment of Wu et al.~\cite{Wu1986} (see Fig.~\ref{first-experiments}c), in 1986 made use of a sub-threshold OPO where a MgO:LiNbO$_3$ crystal was placed in linear optical cavity (see Fig.~\ref{first-experiments} c). A frequency doubled laser beam at 532\,nm served as the pump field to drive the system close to threshold, thereby generating squeezed vacuum in the down-converted field at 1064\,nm. The resulting squeezed vacuum state was analyzed by means of homodyne detection comprising a bright local oscillator (LO) at 1064\,nm, a balanced beam splitter and two photodiodes. The difference of the photocurrents was recorded and fed into a spectrum analyzer that displays the power spectral density corresponding to the variance of the quadratures of the squeezed vacuum. By scanning the phase of the LO, all quadrature variances could be traced out as shown on the graph in Fig.~\ref{first-experiments}c. The dashed line represents a calibration of the vacuum noise limit, and thus squeezed light production was clearly witnessed.    

Polzik et al.\ changed the configuration to a bow-tie shaped ring cavity with a KNbO$_3$ crystal and generated frequency tunable squeezed light of $3.8$\,dB for spectroscopy~\cite{Polzik1992}. Using a MgO:LiNbO$_3$ crystal as a monolithic cavity design where the end-facets of the crystal were curved and coated with high-reflective mirror coatings, Breitenbach et al.\ \cite{Breitenbach1997} obtained $6$\,dB squeezing while Lam et al.\ achieved approximately $7$\,dB vacuum squeezing in an unlocked system~\cite{Lam1999}. Such a monolithic system for non-classical light generation was pioneered by Sizmann et al.~\cite{Sizmann1990,Kurz1992} for the inverse process of up-conversion as discussed below. Stable squeezing of $6$\,dB for hours of operation was achieved by Schneider et al.\ using a semi-monolithic cavity system that was seeded with a displacement beam~\cite{Schneider1998}.  

\begin{figure}
	\centering
		\includegraphics[scale=1]{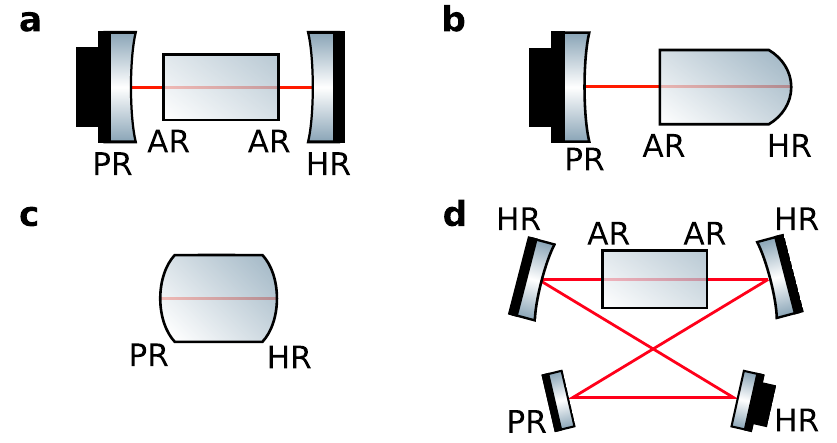}
    \caption{Different cavity geometries used to produce squeezed light by parametric down-conversion. (a) Linear cavity with crystal in the middle, (b) semi-monolithic linear cavity where one end-face of the crystal is curved. (c) monolithic linear cavity formed by the two end-faces of the crystal. (d) Bow-tie travelling-wave cavity. Coatings at the signal and idler wavelengths: AR: anti-reflective coating, HR: high-reflective coating, PR: partially reflective coating.}
	\label{squeezer-geometries}
\end{figure}

For several years, the achievable squeezing degree leveled around $6$\,dB which was caused by the (at that time inevitable) intra-cavity losses, detector losses and phase noise. A turning point for squeezed light generation via OPO occurred around the year 2006 where the $6$\,dB squeezing limit was surpassed first in a ring-cavity configuration with a periodically poled KTiOPO$_4$ (PPKTP) crystal (achieving $7.2$\,dB in 2006~\cite{Suzuki2006} and $9$\,dB in 2007~\cite{Takeno2007}) and later in a linear-cavity configuration using a LiNbO$_3$ crystal (achieving the magic $10$\,dB squeezing)~\cite{Vahlbruch2008}. These experiments have been further optimized and squeezing levels of more than $10$\,dB have been achieved in all cavity configurations shown in Fig.~\ref{squeezer-geometries} except the linear cavity with crystal in the middle (Fig.~\ref{squeezer-geometries}a). Thereby $12.7$\,dB squeezing was measured with a PPKTP crystal using a monolithic cavity~\cite{Eberle2010} (see Fig.~\ref{SqzRecords}a), $12.3$\,dB with a semi-monolithic cavity~\cite{Mehmet2011} and $11.6$\,dB with a ring cavity~\cite{Stefszky2012}. Moreover, by reducing technical noise at very low frequencies, squeezed light has been observed in the audio frequency regime which is important for applications in Gravitational wave detection~\cite{McKenzie2004,Vahlbruch2006}.

The technical advances necessary for reaching high squeezing levels concern all three types of noise mechanisms mentioned above: intra-cavity losses, detection losses and phase noise. Intra-cavity losses were especially reduced by introducing PPKTP crystals which exhibit low optical absorption at the (fundamental) squeezing wavelength~\cite{Steinlechner2013} and a negligible amount of pump-induced absorption (a problem that cannot be neglected in other crystals~\cite{Suzuki2006}). Furthermore, the development of low-loss coatings helped to reduce intra-cavity loss so that escape efficiencies of more than $97$\,\% (defined by the coupling rate divided by the intra-cavity loss rate) have been achieved~\cite{Takeno2007,Vahlbruch2008,Eberle2010}. Detection losses are mainly due to an imperfect quantum efficiency of the photo diodes and the visibility between local oscillator and squeezed beam at the homodyne detector's beam splitter. While photo diodes for visible wavelength based on Silicon with a quantum efficiency close to $100$\,\% were already available for a long time, e.g.~\cite{Polzik1992}, high efficiency InGaAs photo diodes for infrared wavelengths became available only about 8 years ago~\cite{Dong2008}. Visibilities of the interference contrast at the homodyne detector close to $100$\,\% were also very important to measure high squeezing values. Reducing the phase noise was achieved by filtering the pump beam with a ring cavity~\cite{Vahlbruch2008} and by optimized feedback systems for cavity and phase locks~\cite{Takeno2007}.

Squeezing in an OPO cavity was not only achieved below threshold, but also above, separately in both signal and idler beam~\cite{Fuerst2011}. Using a whispering gallery mode resonator (WGM) made of LiNbO$_3$ with a record-low threshold power of only a couple of $\mu W$, $1.2$\,dB squeezing could be observed.  This squeezing in only one of the nondegenerate OPO beams is hardly important for applications but relevant for better understanding the process. In the above-threshold operation, classical noise introduced by the pump is a critical problem for observing squeezed light. When exploiting the combined action of both signal and idler in the degenerate or non-degenerate regime, WGMs are a promising source of sizeable squeezed light usable in real-world applications as they are compact, tunable and operational with low pump powers.

\subsection{Parametric up-conversion}

Parametric up-conversion, where a fundamental input beam undergoes a frequency doubling process, is the inverse process of down-conversion at the doubled frequency. This process has also shown to be a viable strategy for producing squeezed light. The first experiments were carried out by Sizmann et al.\ and by K\"urz et al.~\cite{Sizmann1990,Kurz1992} in a double resonating system where both the fundamental as well as the frequency doubled mode was supported by a solid monolithic cavity.  A single resonant system has been also realized and used for demonstrating squeezing of the frequency doubled mode~\cite{Paschotta1994}. However, since squeezing of this mode for a single resonant system is theoretically limited to a finite value, this strategy has not been followed by many groups.

\subsection{Single-pass optical parametric amplification}

\begin{figure*}
	\centering
		\includegraphics[scale=1]{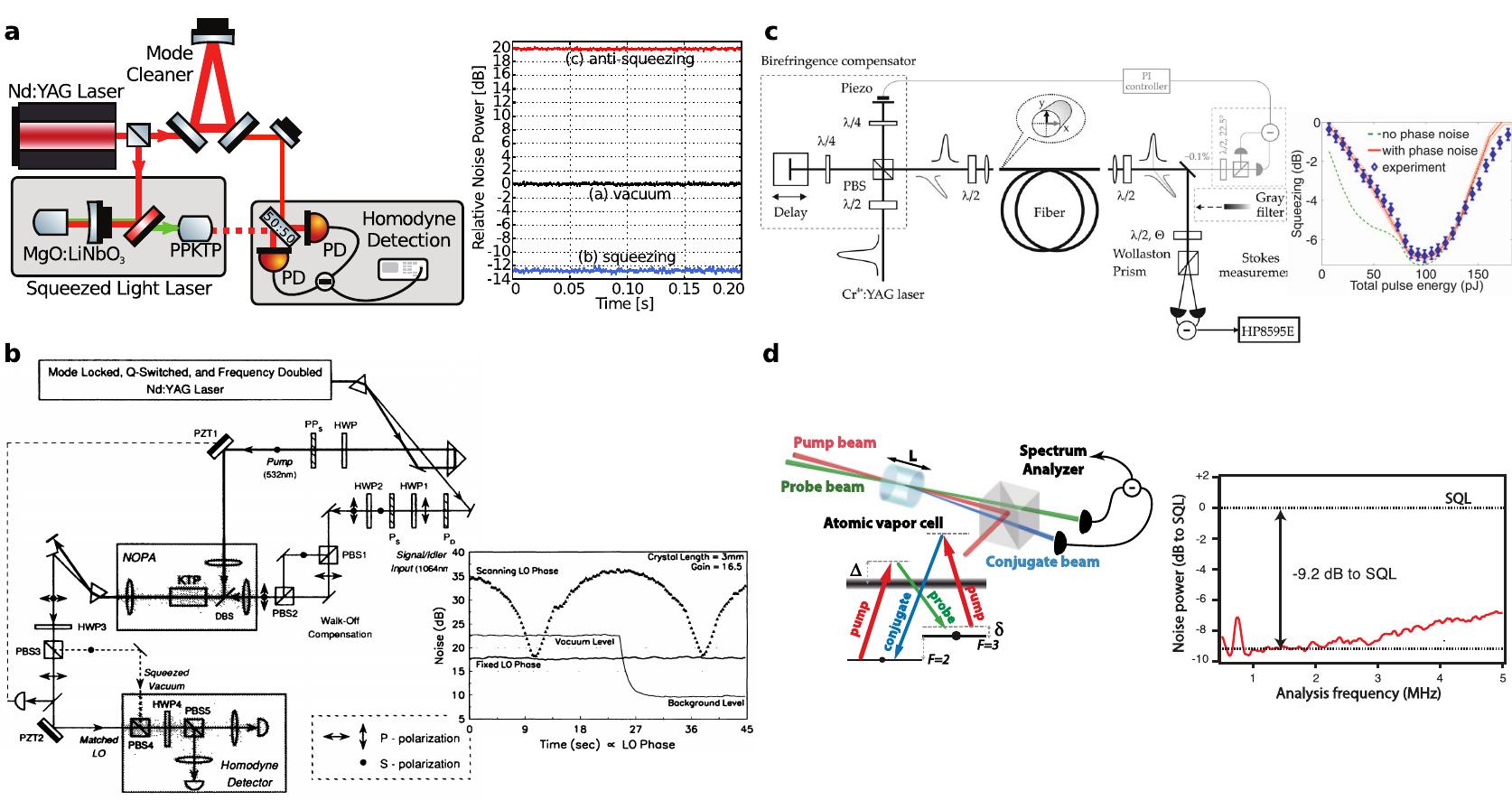}
    \caption{Squeezing records in different systems. (a) 12.7\,dB squeezing was generated by parametric down-conversion in a monolithic PPKTP cavity using a frequency-doubled 1064\,nm continuous-wave laser beam as pump. The modecleaner improves the spatial beam profile of the local oscillator and thus enhances the interference contrast in the homodyne detector.~\cite{Eberle2010} (b) A single-pass non-degenerate OPA (NOPA) consisting of a KTP crystal pumped with a frequency-doubled, mode-locked, q-switched laser was used to generate 5.8\,dB squeezing. The process was Type II phase matched which means that bright squeezing was produced in two orthogonal polarization modes. This was used to form a squeezed vacuum state and a bright local oscillator by means of a balanced beam splitter~\cite{Kim1994}. (c) 6.8\,dB polarization squeezing was generated by the optical Kerr effect in a polarization maintaining fiber using a pulsed 1500\,nm laser. The squeezing was measured with a Stokes parameter measurement scheme. To compensate for the birefringence of the fiber, the two orthogonal pulses were shifted in time prior to fiber injection~\cite{Dong2008}. (d) A double lambda scheme is employed to generate strongly correlated beams via four-wave-mixing. A pump and a probe beam interacts in a cloud of Rubidium atoms and subsequently forms correlations between the probe and its conjugate beam. The results are recorded on a spectrum analyzer (SA) and a maximum of $9.2$\,dB squeezing was measured~\cite{Glorieux2011}.} 
    \label{SqzRecords}
\end{figure*}

An alternative way of addressing the small $\chi^{(2)}$ non-linearity is to significantly increase the pump amplitude by using ultra-short pump pulses. This will effectively increase the non-linearity and thus the production of squeezed light. However, the pulse shape also leads to complications as the amount of squeezing strongly depends on the temporal mode shape overlap of the pump and the LO pulse. Despite of these complications related to the pulsing operation, squeezing from a pulsed parametric down-conversion source was observed already in 1987 by Grangier et al.~\cite{Slusher1987}, and significantly improved in Refs.~\cite{Aytur1992} and \cite{Kim1994} in the group of Kumar. In the latter work (shown in Fig.~\ref{SqzRecords}b), a mode-tailored local oscillator was produced by injecting an auxiliary beam into the squeezing crystal in a polarization mode orthogonal to the signal mode, but still actively amplified using a KTP Type II phase matched crystal. This resulted in a spatio-temporal mode of the LO near identical to that of the squeezed signal mode. Due to this mode optimization strategy of the LO, a record-breaking squeezing value of 5.8\,dB was measured in 1994. After more than 20 years of research, this is still the quadrature squeezing record in pulsed optical parametric amplification systems. Pulsed squeezed light experiments based on optical parametric amplification have been extended to include periodically poled crystals~\cite{Hirano2005}. 

By confining the pump field in the nonlinear crystal over a long distance by means of a crystalline waveguide structure (thereby circumventing diverging beams due to diffraction), it is possible to increase the effective nonlinearity even further. Using such systems single-pass pulsed squeezing~\cite{Eto2007,Hirosawa2009,Zhang2007} and even single-pass continuous wave squeezing has been observed~\cite{Yoshino2007,Pysher2009}.

\section{Squeezed light from optical fibers}
As an alternative to second order nonlinear processes also third order nonlinearities can serve to generate squeezed light. Third order nonlinear processes are much weaker than their second order counterparts, however they also exist in amorphous materials, rendering the choice of the physical system more flexible.  The weak interaction can successfully be compensated by a long interaction length. This is why squeezing inside optical fibers is a practical alternative. Optical fibers also offer low loss over long distances. One of the highly active competitors to the first squeezing experiment was in fact a setup using an optical fiber. At IBM Shelby et al., \cite{Shelby1986}, used a $114$\,m long optical fiber and injected high power continuous wave laser beams. In hind side this experiment can be regarded as quite heroic as several obstacles that nowadays are considered interesting physics in their own right had to be identified and overcome (see introduction).

\begin{figure*}
	\centering
		\includegraphics[scale=1]{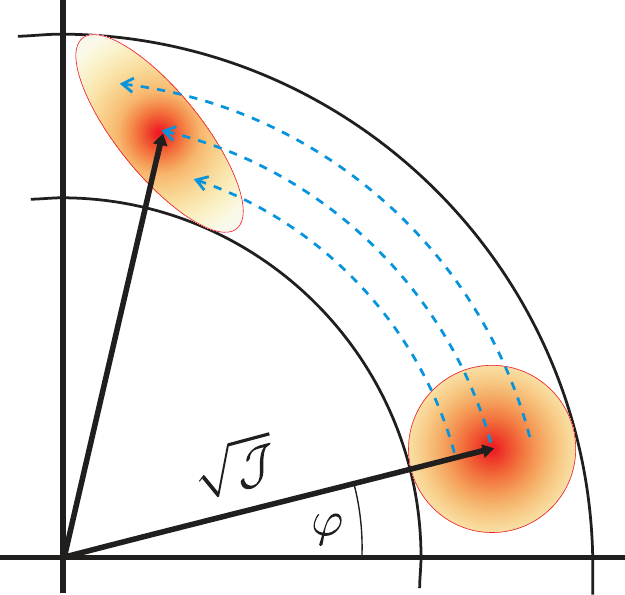}
		\caption{Squeezing of a coherent state by the nonlinear optical Kerr effect in quadrature phase space. The acquired phase shift varies with amplitude $\sqrt{I}$ and transforms the quantum state uncertainty into an ellipse~\cite{Rigas2013}.}
	\label{Kerr-squeezing}
\end{figure*}

Squeezing in optical fibers relies on four-wave-mixing and the nonlinear optical Kerr effect. Third order nonlinearities lead to a situation where the refractive index of the material in which the light is propagating depends on the intensity of the light itself:
\begin{equation}
n = n_0 +n_2 \cdot I\ .
\end{equation}
The Kerr effect transforms a coherent state from a pump laser into a squeezed state (see Fig.~\ref{Kerr-squeezing}). This is best explained by regarding the acquired phase shifts in phase space. Regions in phase space with higher amplitude are associated with an increased phase shift as a direct consequence of the nonlinear refractive index. Hence the circular uncertainty region of the coherent state gets transformed into a squeezed state (this is true in a first order approximation which practically can be applied as high intensities are needed for a considerable phase shift~\footnote{The proper quantum treatment will lead to a periodic evolution giving rise to cat states. But the loss in a typical fiber is not low enough, or rather the nonlinearity is not high enough to reach this regime~\cite{Sanders1992}}).
The nonlinear Kerr effect does not need sophisticated phase matching strategies and thus is one of the simplest means to produce squeezed light.
The challenges here stem from the fact that third order nonlinearities are weak and high power levels are needed that also can trigger other unwanted nonlinear interactions.

The first experiment in 1986 (see Fig.~\ref{first-experiments}b) used a continuous-wave pump to generate the squeezing. As the required pump powers are significant it was soon realized that this leads to a number of different effects inside the fiber that hinders or reduces the generation of squeezing. The most important negative effect here is photon-phonon coupling inside the fiber.  Guided Acoustic Wave Brillouin scattering \cite{Levenson1985} introduces noise at acoustic frequencies up to the GHz regime, thus the fiber has to be cooled and these frequencies have to be avoided in measurements. At the required power levels stimulated Brillouin scattering backscattered most of the launched light. To stay below the Brillouin threshold a phase modulation scheme generating several different wavelengths had to be used. The measurement itself cannot easily rely on homodyne technology as the intensities involved are already very high. Thus it is technically not possible to employ an even brighter LO reference beam. Instead Shelby et al.\ used a phase shifting cavity~\cite{Galatola1991,Villar2008}. By reflecting off a cavity resonance the bright carrier can be phase shifted compared to the radio frequency sidebands. A detuning from the cavity enables to measure a rotated projection of the squeezing ellipse in phase space.

Later experiments could lower the technical difficulties significantly by using short pulses \cite{Slusher1987,Bergman1991,Rosenbluh1991}. The short time scales and high peak intensities of these pulsed systems reduce the average power required into regimes that are more favorable for the detection system and do not suffer from stimulated Brillouin backscattering. In order to avoid dispersive pulse spreading in the optical fibers one can go into a regime of optical solitons \cite{Rosenbluh1991, Drummond1993}. In addition, in the soliton regime the correlations between wavelength components can be used to generate squeezed beams by spectral filtering \cite{Friberg1996, Spaelter1998}.

The Kerr effect is photon-number preserving and the squeezing is thus not measurable in direct detection as noted above. Measuring other quadratures by homodyne techniques often suffers from the signal beam being already very intense. As a viable alternative to spectral filtering (see above) one may utilize interferometric techniques to be able to measure the squeezing~\cite{Kitagawa1986}. 

One interferometric solution is to use a cavity as presented above. Another successful solution uses a Sagnac interferometer employing either a balanced or unbalanced beam splitter. Using the balanced beam splitter the two counter propagating beams get squeezed. At the output both beams interfere destructively resulting in a vacuum squeezed beam and constructively resulting in a bright beam that can be used as a local oscillator \cite{Bergman1991}. With an unbalanced beam splitter one of the counter propagating beams is bright and gets squeezed. The other (weak) beam interferes at the output and shifts the squeezed beam in phase space. For certain power levels this interference leads to directly measurable amplitude squeezing \cite{Werner1997,Schmitt1998,Krylov1998}. To introduce more flexibility into the setup that allows for independent control of the relative intensities and phase of the two pulses, a linear version of the asymmetric Sagnac interferometer was suggested and demonstrated~\cite{Fiorentino2001}.  

Interferometric setups can be largely simplified when employing polarization.
Polarization squeezing denotes the reduction of the quantum uncertainty of the polarization of the light field. The concept of polarization squeezing relies on a Heisenberg uncertainty relation in polarization variables that uses Stokes operators and is close to what one uses to describe the uncertainty of the atomic spin. For intense beams (as used in Kerr squeezing) polarization squeezed beams can be well approximated by quadrature variables.

Polarization squeezing can be achieved by a single pass of optical pump pulses on the two polarization axes of a polarization maintaining optical fiber. When compensating for the birefringence inside the fiber the two orthogonal polarized squeezed beams interfere at the output of the fiber. The resulting polarization squeezing can then be determined by a Stokes measurement. The simplicity of the setup and very good spatial and spectral overlap of the two interfering beams led to a measured squeezing of around $7$\,dB \cite{Heersink2005,Dong2008} (see Fig.~\ref{SqzRecords}c). A detailed theoretical analysis shows that indeed Brillouin and Raman scattering are limiting the squeezing in practice \cite{Corney2008}.

Photonic crystal fibers (PCFs) offer new possibilities in generating squeezed states.  Here the light is guided by an effective refractive index through a microstructure around a solid core or a photonic crystal band gap that enables guiding inside a hollow core. These type of fibers offer much higher effective nonlinearities and flexibility in dispersion management. Squeezing in PCFs has been shown early after their development \cite{Lorenz2001, Fiorentino2002, Hirosawa2005}. The involved powers are lower than those needed in standard fibers and wavelengths outside the telecom band can be utilized. The strong nonlinearities and dispersive properties of PCFs however require a careful balancing and so far the best squeezing results still have been achieved with standard fibers.

\section{Squeezed light from atomic ensembles}
As mentioned in the introduction, the first evidence of squeezing the optical field was obtained through the non-linear interaction between light and atomic vapor. The motivation for using an atomic ensemble for squeezed light generation at the early days of quantum optics was that the intrinsic nonlinearity associated with light-atom interaction can be very large -- significantly larger than using $\chi^{(2)}$ nonlinear crystals. The simplest kind of a nonlinearity is atomic saturation which is observed when the ensemble is illuminated by laser light resonant with the atomic transition. However, it is a four-wave mixing mechanism that leads to the formation of squeezed light. The process is enabled by a $\Lambda$-shaped atomic energy level configuration where two ground states are coupled to a single excited state with optical pump beams that are either frequency degenerate or non-degenerate. The pump beams enable a cycling from one ground state to the other and back under the emission of two photons that are frequency non-degenerate by an amount given by the frequency difference of the two ground state levels. These two modes (occupied by the photons) are known as Stokes and Anti-Stokes and they form a two-mode squeezed state in which squeezing can be observed by using a local oscillator of frequency located between the two Stokes modes. 

In the experiment of Slusher et al.~\cite{Slusher1985} (see Fig.~\ref{first-experiments}a) the nonlinear interaction took place between laser light and a beam of Sodium atoms. To enhance the effect, the atomic beam was embedded in two cavities  resonant for the Stokes and Anti-Stokes sidebands. As a result of the non-linear interaction, correlations between the two sidebands were established and thus squeezed light could be measured. The results of a homodyne measurement are shown in Fig.~\ref{first-experiments}a. A modest squeezing degree of $0.3$\,dB was observed in this experiment. 

This small amount of squeezing measured in this initial experiment was mainly due to other detrimental non-linear processes such as Raman scattering and fluorescense occurring simultaneously with the four-wave-mixing process. These processes lead to incoherent emission of noisy modes into the squeezed modes and thereby a degradation of the squeezed state. These effects have been limiting the amount of squeezing in atomic systems for a number of years. However, by considering intensity-difference squeezing (also known as twin beam squeezing, see below) of a double-$\Lambda$ system, many of the parasitic processes cancel out, and thus strong squeezing can be revealed. This led the group of Paul Lett to observe close to 9\,dB squeezing in the intensity difference of the two output modes in four-wave-mixing in atomic vapor of Rubidium~\cite{McCormick2007,McCormick2008}. A pump and a probe beam are incident onto the atomic ensemble, and interact efficiently via the four-wave-mixing process. The probe and the conjugate are amplified in the process establishing quantum correlations that are revealed through intensity-difference detection. The noise of the intensity difference was squeezed by $8.8$\,dB below shot noise level. Using improved techniques, the intensity-difference squeezing was increased to 9.2dB which is the squeezing record for squeezing in atomic ensembles~\cite{Glorieux2011}. This experiment is shown in Fig.~\ref{SqzRecords}d.  
A similar system has been used also to produce single-mode squeezing with a squeezing degree of $3$\,dB~\cite{Corzo2011}.   

It is also possible to observe squeezing through a Faraday rotation nonlinearity in atomic systems~\cite{Ries2003,Barreiro2011}. Here a polarization mode orthogonal to the polarization of the pump is squeezed as a result of a cross-phase modulation nonlinearity. Using this atomic nonlinerity, $0.9$\,dB~\cite{Ries2003} and $2.9$\,dB~\cite{Barreiro2011} squeezing have been measured.    

The benefits of a fiber (strong confinement over a long distance) and atomic vapor (high nonlinearity) can be merged in a single system by filling a hollow core PCF with gaseous or liquid materials. This offers promising new possibilities of efficient nonlinear interaction for the generation of squeezing largely avoiding Brioullin and Raman scattering. First results have demonstrated squeezing in fibers filled with high pressure gas \cite{Finger2015} and atomic vapour \cite{Vogl2015}.

\section{Squeezed light from semiconductor lasers}
Going back to the days of vacuum tubes, it was a major invention that current controlling or amplifying vacuum tubes can work in a sub shot noise regime provided that the current is in the space charge limited regime \cite{Schottky1937}. This leads to electrons arriving more evenly spaced than expected for shot noise. If one only could transfer each of these electrons into one photon, then one would have sub shot noise light. An obvious candidate for such a transfer is a light emitting diode or a light emitting laser diode. The latter type is preferable if one wants to ultimately collimate the light. Machida, Yamamoto and Itaya followed this approach and in 1986 succeeded in measuring squeezed light emitted directly from a laser diode, the noise reduction being 0.33 dB \cite{Machida1987}. In the following years the same group tried to improve the set-up and to optimize the parameters of operation to minimize the noise. Eventually they managed to observe a very large noise reduction by positioning the detector and the laser face to face \cite{Richardson1991}. Other groups who first collimated the light beam before detecting the squeezing could at first not come any where near reaching similar numbers. The noise reduction in the collimated light emission from laser diodes lies around 3-4 dB \cite{Bramati1997}. Ultimately it became clear that in order to observe larger noise reduction, one has to measure the whole emission, as might be guessed from the argument above, i.e. one has to measure all modes. There are strong correlations between different spatial modes \cite{Poizat1998}. Some of these modes with negligible power help to significantly reduce the noise (for details see \cite{Bramati1997} and references there in).

\section{A short note on multimode squeezing}

The field of multimode squeezing and entanglement is a highly topical field due to the associated interesting applications in quantum communication, quantum computing and quantum sensing. Multimode squeezing will be briefly discussed in this section. Although we have mostly been referring to ``single mode'' squeezing in previous sections, as a matter of fact, all squeezed states are multimode states possessing multi-mode quantum correlations. Squeezing is a result of quantum correlations or entanglement between sideband frequency modes. As a result, their joint measurement with a single local oscillator (located in between the sidebands) will reveal the so-called single-mode squeezing. This was experimentally addressed in Ref.~\cite{Huntington2005} where the entanglement of the two sidebands was directly measured. When the sidebands are ``widely'' separated, either in distinct frequency or polarization modes, the modes become easier to access. In some cases, only the intensity correlations can easily be measured due to complications in performing homodyne detection. For such measurements, only squeezing between a single pair of quadratures have been witnessed and this is often referred to as twin-beam squeezing. Such an experiment was demonstrated for the first time already in 1987 in an OPO operating above threshold~\cite{Heidmann1987} holding the world record in squeezing for many years. Later, the same effect has been realized in a number of other systems including seeded OPO~\cite{Zhang2000}, new versions of the above-threshold OPO (in a bulk cavity~\cite{Laurat2003,Fuerst2011} and a micro cavity~\cite{Dutt2015}), seeded~\cite{Aytur1990} and unseeded single-pass OPA~\cite{Iskhakov2009}, waveguide OPA~\cite{Eckstein2011} as well as fiber~\cite{Guo2012} and atomic based systems~\cite{McCormick2007}. There are also a number of recent studies on squeezing and entanglement between a large number of different modes which includes standard two-mode squeezing~\cite{Ou1992,Reid1989}, N-mode entanglement~\cite{Aoki2003,Jing2003,Su2007,Yukawa2008,Su2012}, entanglement between spatial modes~\cite{Choi1999,Boyer2008,Lassen2009,Janousek2009}, frequency modes~\cite{Roslund2013,Chen2014,Gerke2015}, temporal modes~\cite{Yokoyama2013}, mixed modes~\cite{Gabriel2011,Liu2014} and different colors~\cite{Coelho2009}. Using these multimode squeezing processes it is in principle possible to generate large cluster states that are the main resource in linear quantum computing~\cite{Menicucci2006}, or to produce spatial correlations that can be used for squeezing-enhanced imaging~\cite{Kolobov2000}.

\section{Final remarks}
Squeezed light has come a long way since its first demonstration 30 years ago. Significant advancements have been made from the initial $0.3$\,dB squeezing till todays near $13$\,dB squeezing. It is, however, interesting to note that the experimental platforms of nonlinear crystals, fibers and atomic ensembles used in 1985 and 1986 are the same as those used today for generating highly efficient squeezing. The advancements have mainly been of technical nature, that is, successful development of low-noise electronics for phase locking, low loss optical components and high efficiency photo diodes have led to largely improved systems. 

In addition to these technical advancements, within the last few years there have been demonstrations of the production of squeezed light in new systems. Using single emitters such as a single ion in a high finesse cavity~\cite{Ourjoumtsev2011} or a single semiconductor quantum dot~\cite{Schulte2015} small degrees of squeezed light has been generated. In these accounts, the intrinsic strong nonlinearity between a two-level system and light was employed. Recently, it was also shown that squeezed light can be produced on a micron-sized platform exploiting the nonlinear coupling between light and a mechanical oscillator. Here a suspended mechanical oscillator is embedded into a cavity and displaced by the radiation pressure force produced by the interacting circulating light. The displacement of the mechanical oscillator will in turn shift the phase of the circulating field. In essence, this corresponds to an intensity dependent phase shift which is reminiscence of the optical Kerr effect that is known to squeeze the field. Experiments have been performed in a photonic crystal cavity supporting phononic and photonic modes simultaneously~\cite{Safavi-Naeini2013}, and in a bulk cavity setup containing a mechanical membrane~\cite{Purdy2013}. Another route to squeezed light generation on micro-platforms is to exploit the Kerr non-linearity of CMOS compatible materials such as Silicon Nitride (SiN). E.g. there has been recent reports on intensity difference squeezing~\cite{Dutt2015} and single mode squeezing~\cite{Iskhakov2016} generated in SiN micro-ring cavities. All these recent demonstrations of squeezed light in completely new and miniaturized settings might also represent a guide towards the future trends in squeezed light generation: Generating squeezing on smaller and potentially more rugged platforms allow for up-scaling and eventually real-life applications. 

Initially, squeezed light was envisaged to enable enhanced communication rates~\cite{Yuen1978,Shapiro1979,Braunstein2000,Jing2003} and improved detection of weak forces such as gravitational waves~\cite{Caves1981,Leuchs2002,Chua2014}~\footnote{The high precision measurement of many quantities can be traced back to a position measurement. For a long time the precision with which the position of a massive particle can be determined was thought to be limited by Heisenberg's uncertainty relation, where the uncertainty is minimized and distributed equally between position and momentum. The term standard quantum limit (SQL) was coined to refer to this situation. It is enlightening to follow the development of how the scientific community came to understand that it is possible to go well below this SQL\@. Braginsky pioneered the the concept of quantum non-demolition measurements developing mechanical resonance detectors for gravitational waves~\cite{Braginsky1980}. A different approach applied to optical interferometers was proposed by Unruh showing that an interferometer can become more sensitive~\cite{Unruh1983}. But he used his own notation and the community somehow did not really catch on. At about the same time Yuen pointed out that the trick is to correlate the uncertainties in position and momentum in a particular way~\cite{Yuen1983}. He used the term contractive states but the concept seemed to be academic with little relevance to applications. It took another 10 years until Jaekel et al.~\cite{Jaekel1990} treated the problem using the linearization approach and soon after Luis and Sanchez-Soto~\cite{Luis1992}, providing a full quantum treatment in the language of quantum optics. The community now fully appreciated the relevance of correlated uncertainties, allowing for improving the precision from the SQL (proportional to one over the square root of the photon number) to the much lower Heisenberg limit (proportional to one over the square root of the photon number).  At this stage it was clear how optics can help: increase the power of the laser feeding the interferometer until the SQL is reached. Then squeeze the field at the second, dark input port of the interferometer and tilt the squeezed ellipse in phase space relative to the laser light at the bright input port to maximally anti-correlate the in-phase and the out-of-phase quadrature uncertainties -- done. Ideally the squeezing angle is noise frequency dependent to allow for broad band detection with better than SQL precision~\cite{Kimble2001}. However, all is easier said than done and the experimental demonstration is still at stake.}. The latter was demonstrated first at the GEO600 gravitational wave detector~\cite{LSC2011} and later at the LIGO detector~\cite{Aasi2013}.  
These applications proposed more than 35 years ago are still some of the most prominent applications of squeezed light. However, in addition to these applications, squeezed states have also been shown to be the resource of quantum teleportation~\cite{Braunstein1998,Furusawa1998}, continuous variable quantum computing~\cite{Menicucci2006}, quantum error correction coding~\cite{Aoki2009,Lassen2010}, phase estimation~\cite{Berni2015} and tracking~\cite{Yonezawa2012}, fundamental tests of quantum mechanics (such as the Einstein-Podolsky-Rosen gedanken experiment)~\cite{Reid1988,Ou1992,Reid2009}, quantum imaging~\cite{Kolobov2000, Treps2002} of e.g. biological samples~\cite{Taylor2013}, clock synchronization~\cite{Giovannetti2001} and magnetometry~\cite{Wolfgramm2010,Otterstrom2014}. Moreover, in recent years, a squeezed light source has been the working horse for quantum state engineering, in particular non-Gaussian state generation using the method of photon subtraction~\cite{Ourjoumtsev2006,Neergaard-Nielsen2006,DellAnno2006,Andersen2015} as required for various quantum processing protocols~\cite{Eisert2002,Fiurasek2002,Giedke2002,Niset2009,Bartlett2002}. Thus, a plethora of new and exciting applications of squeezed light have appeared since the initial proposals, and it will be exciting to observe where these application studies will take us the next 30 years.

\section{Acknowledgements}
We acknowledge support from the Lundbeck Foundation, the Danish Council for Independent Research(Sapere Aude 0602-01686B and 4184-00338B) and the H.C. \O rsted postdoc programme.

\bibliographystyle{aipnum4-1}
\bibliography{bibliography}{}

\end{document}